\documentstyle[twoside,fleqn,espcrc2]{article}
\newcommand{\half}{{\textstyle{1\over 2}}}
\newcommand{\quart}{{\textstyle{1\over 4}}}  
\newcommand{\Tr}{\hbox{Tr}}

\newcommand{\Complexes}{{\rm C}\llap{\vrule height6.3pt width1pt
depth-.4pt\phantom t}} 
\input epsf.tex
\title{Quantum Gravity and  Regge Calculus}
\author{ Giorgio Immirzi
\address{ Dipartimento di Fisica and I.N.F.N., Perugia;
\\ immirzi@perugia.infn.it \\ II meeting
on Constrained Dynamics and Quantum Gravity, Santa Margherita Ligure,
Sept.96.\\ For Tullio Regge on his 65th birthday}}

\begin{document} 
\begin{abstract} This is an informal review of the
formulation of canonical general relativity and of its implications
for quantum gravity; the various versions are compared, both in the
continuum and in a discretized approximation suggested by Regge
calculus. I also show that the weakness  of the link with the
geometric content of the  theory  gives rise to what I think is a
serious flaw in the claimed derivation of a discrete structure for
space at the quantum level. 
\end{abstract} 
\maketitle
\section{Introduction} 
From chromodynamics, I have inherited  the
prejudice that a decent quantum field theory must have a credible
discrete approximation. Being deep rooted, this prejudice resists
counterexamples. In particular, Regge calculus \cite{Regge}  makes the
geometric content of General Relativity so clear that I find difficult
to believe that a successful quantization scheme would not  originate,
or translate cleanly into this discretized approximation. And
therefore,  I am reluctant to add  the quantization of Regge calculus
to the list of counterexamples, in spite of the modest success which
the various attempts have achieved \cite{RS,RL,GI,JZ}, and of the
unquestionable fact that what progress there has been has been
achieved ignoring this point of view.

I shall give a quick sketch  of the recent work on  canonical
quantization of gravity, and a tentative critique of this work in the
light of what I claim Regge calculus would suggest, with side comments
on the meaning of the results so far obtained. So, in spite of the
rather generic title, I shall mention only a small fraction of the
work going on.

The idea that gravity should be treated as a connection theory
underlies  all the work I shall be considering, and this is where I
shall begin; the other main ingredient is the idea of the "loop
representation" of Rovelli and Smolin, which has evolved into a theory
of "Spin nets", and this will be discussed next, with a brief summary
of the striking and very recent progress that has taken place. Here I
shall try and explain why I think that these results are spoilt by the
imperfect link of the theory with the geometric content of general
relativity. The crucial importance of this link will be illustrated
using  Regge calculus as an example.

\section{Connections and constraints.}

If the aim is to formulate canonical general relativity as a some sort
of gauge theory on 3--space, we have various options\cite{AII}.

Option (1) is A.D.M. geometrodynamics rewritten as a connection
theory.

\noindent One follows A.D.M. taking space slices $\Sigma_t:\
t(x)=$const., with a congruence $t^a: \ t^a\partial_at=1$  related to
the unit normal to the slice $n^a$ by $t^a=Nn^a+N^a$, or
$n_a=-N\partial_at$; however A.D.M. start with the Einstein--Hilbert
action $S={1\over 2\kappa}\int\sqrt{-g}\;R\;d^4x$, while here  one
starts with an action $S={1\over 4\kappa}\int\epsilon_{IJKL}e^I\wedge
e^J\wedge R^{KL}$ (where $ R^{KL}$ is the curvature 2--form of the
Levi--Civita connection $\Omega^{IJ}$, and $\kappa :=(8\pi
G_{Newton})/c^3$),  and the further choice is made to partially fix
the $O(3,1)$ gauge freedom setting: 
\begin{equation}
e^0_a=-n_a=N\partial_a t  \label{ci} 
\end{equation} 
This choice ("time
gauge") leaves invariance under local $O(3)$ transformations, with
$e^{ia},\ i=1,2,3$  space--like to provide a local frame on the slice.
The pull--back of $\Omega^{IJ}$ to the space slice gives the 3--d L.C.
connection $\Gamma^i_a:=\half q_a^{\ b}\epsilon_{jik}\Omega^{jk}_b$,
and the extrinsic curvature $K^i_a:= q_a^{\
b}\Omega_b^{0i}=e^{ib}K_{ab}$. As canonical variables we take the
pair: 
\begin{equation} 
( E^{ia}:=
\half\epsilon^{abc}\epsilon_{ijk}e^j_be^k_c= \det(e^i_a)\; e^{ia}\  ,\
 K^i_a) \label{cii} 
 \end{equation} 
 They form a canonical pair, just
like the $(q_{ab},\pi^{ab}:=\sqrt{q}(K^{ab}-q^{ab}K) $ variables of
A.D.M., because $E^{ia} E^{ib} = q\;q^{ab}$ implies: 
\begin{equation}
K^i_a \; \delta E^{ia} = \frac{1}{2\sqrt{q}} K_{ab} \delta
(E^{ia}E^{ib})  = -\frac{1}{2} \pi^{ab}\;\delta q_{ab} 
\end{equation}

The $\Gamma^i_a$ have curvature $R^i_{ab}$, and can be expressed in
terms of $E^{ia}$ (and its inverse) solving the 9 linear equations
they satisfy: \begin{equation} D_aE^{ia}=0\quad ;\quad
\epsilon_{ijk}E^{a}_iE^{(b}_jD_aE^{c)}_k=0 \label{civ} \end{equation}
We need to impose the A.D.M. constraint, and to constrain $K^i_a$ to
make sure that $K_{ab}={1\over\sqrt{q}}\,q_{bc}\,E^c_iK^i_a $  is
symmetrical; altogether: 
\begin{eqnarray*} 
{\cal
G}_i:=\epsilon_{ijk}K^j_aE^{ka}\approx0\quad\quad\quad\quad\quad
\quad\quad &&\\ {\cal V}_c:=q_{bc}\nabla_a\pi^{ab}=2E^a_i
D_{[a}K^i_{c]}\approx 0 \quad\quad\quad\quad&&\\ {\cal
H}:={1\over\sqrt{q}}\big(\pi^{ab}\pi_{ab}-\half
\pi^2-q\;R\big)=\quad\quad\quad\quad &&\\ {1\over\sqrt{\det E}}
(2E^{[a}_iE^{b]}_jK^j_aK^i_b+
\epsilon_{ijk}E^{ia}E^{jb}R^i_{ab})\approx 0 && 
\end{eqnarray*}
Everything can be rewitten in terms of $2\times 2$ matrices saturating
$i,\ j,\ldots$ indices with $\tau_i:=\sigma_i/(2i)$; so for ex. the
effect of local rotations becomes: 
\begin{eqnarray}
E^a:=\tau_iE^{ia}&\to &U\,E^a\;U^{-1} \nonumber\\ \Gamma_a:=
\tau_i\Gamma^i_a &\to& U\,(\Gamma_a+\partial_a)U^{-1} \label{cvi}
\end{eqnarray} 
There is nothing particularly new about any of this; in
particular $K^i_a$ fits into the scheme poorly, like an external
field, and the connection is a derived quantity.

Option (2) is a shrewd variation on the theme devised by J. F.
Barbero\cite{Ba,AII}, in a re--examination of the Ashtekar\cite{Ai}
formulation.

For some $\beta$ to be fixed, we can change our basic variables as
follows: 
\begin{equation} (E^{ia},K^i_a)  \;\to\;(E^{ia}\ ,\
A^{(\beta)i}_a:=\Gamma^i_a\,+\,\beta\, K^i_a )   \label{cvii}
\end{equation} 
which {\it is} a canonical transformation, because
\begin{equation} 
E^{ia}\;\delta A^{(\beta)i}_a=\beta E^{ia}\;\delta
K^i_a+{1\over 2}\partial_a(\tilde\epsilon^{abc} e^k_b\;\delta e^k_c )
\end{equation} 
so that: 
\begin{equation} 
\{ A^{(\beta)i}_a(x),E^{jb}(y)\} =\beta\kappa\,
\delta^i_j\delta^b_a\delta(x,y) \label{cix} 
\end{equation} 
and all
other Poisson brackets vanish (in particular, $\{
A^{(\beta)i}_a,A^{(\beta)j}_b\}=0$). However unlike $K^i_a$, 
$A^{(\beta)i}_a$ transforms like  an $SU(2)$ connection for any
$\beta$, and one can introduce $D^{(\beta )}_a$ derivatives and  a
curvature $F^{(\beta )i}_{ab}$. Using the properties of the
Levi--Civita connection and a bit of algebra the constraints become:
\begin{eqnarray*} 
D^{(\beta )}_aE^{ia} =\beta{\cal G}_i\ \approx\ 0
\quad\quad\quad\quad&&  \\ E^{ia}F^{(\beta )i}_{ab}=\beta\;{\cal
V}_b+\beta^2\;K^j_b{\cal G}_j\ \approx\ 0 \quad\quad\quad &&\\ {\cal
H}= {1\over\sqrt{\det (E)}}\epsilon_{ijk}E^{ia}E^{jb}F^{(\beta
)k}_{ab} \quad\quad\quad\quad\quad &&\\ - 2{(1+\beta^2)\over\sqrt{\det
(E)}} E^{[a}_iE^{b]}_jK^i_aK^j_b +\ldots\ \approx\ 0  &&
\end{eqnarray*} 
where the dots are terms proportional to $\beta{\cal
G}_i$ and its derivatives.

Clearly the first two constraints are in an acceptable form, and the
last one is not: the bit proportional to $(1+\beta^2)$ is a mess, the
overall factor $1/\sqrt{\det (E)}$ is unpleasant. As for this last
point: we either learn to live with the factor, or we ignore it, 
absorbing it in the Lagrange multiplier. The messy bit is more
troublesome. The quickest way to get rid of it would be to take
$\beta=i$. This is Ashtekar's original choice (together with the idea
of absorbing the factor $1/\sqrt{\det (E)}$ in the Lagrange
multiplier)\cite{Ai} . The point is that there is much more to be said
for this choice: it is (or, one can convince oneself that it is) the
one choice that is geometrically and physically well motivated. In
fact the connection $A^{(i)i}_a$  is the pull--back to the space slice
of the self--dual connection, and $E^{ia}$ the pull--back of the
self--dual product of two vierbeins: 
\begin{eqnarray}
A^{(i)i}_a&:=&q_a^{\ b}C^i_{IJ}\Omega_b^{IJ}:=q_a^{\ b}(
-\half\epsilon_{ijk}\Omega^{jk}_b+i\Omega^{0i}_b)\nonumber\\
E^{ia}&:=& - \epsilon^{abc}C^i_{IJ}e^I_be^J_c \label{cxi}
\end{eqnarray}

This definition of $E^{ia}$ coincides with eq.~(\ref{cii}) if we
assume "time gauge", eq.~(\ref{ci}); for the definition of
$A^{(i)i}_a$ to coincide with eq.~(\ref{cvii}) we have to assume "time
gauge" and to impose explicitly that the connection $\Omega^{IJ}_a$ is
Levi--Civita, which follows if we add to the constraints the {\it
real} part of the second eq.~(\ref{civ}) as a brand new  "reality"
condition. The result is, I think, physically splendid, because we
retain the full Lorentz group as gauge group of the canonical theory,
but  technically appalling, because of the obvious complications
implicit in the use of complex variables. In particular, the $i$ which
now occurs in the Poisson brackets eq.~(\ref{cix}) puts all simple
minded quantizations  on collision course with the desire to have the
operators corrsponding to $E^{ia}$ hermitean. More particularly, I
have not found a way round this difficulty, simple minded or not.

The alternative that has become most popular is to take $\beta=1$, the
"Barbero connection".  I would like to claim that the trouble with
this choice is that the connection has no obvious geometric meaning,
and that there is nothing special about the value 1. In fact, $\beta$
can be anything, and  I shall leave it arbitrary (but real $>0$) in
the following. In Euclidean (++++) gravity the second term of $\cal H$
is proportional to $(1-\beta^2)$, and the choice $\beta =1$ {\it is}
natural, just like $\beta =i$ was "natural" for the (-+++) signature,
but that's not saying much, unless somehow we learn to "Wick--rotate"
the theory. For this possibility \cite{AII}  see later.

There are at least two smart ways out of the inconvenience that
$K^i_a$ is a complicated function of $A^{(\beta )i}_a$ and $E^{ia}$
(so that the expression of $\cal H$ is messy), which  come by
roundabout arguments devised by T. Thiemann\cite{Tb}. I begin by
listing some identities; let $f(x),\ h(x)$ be nice test functions on
$\Sigma$, and 
\begin{eqnarray} {\cal
H}^E[f]&:=&\int_\Sigma{f\over\sqrt{\det E}} \epsilon_{ijk}E^{ia}E^{jb}
F^{(\beta )k}_{ab}d^3x\nonumber\\ V[h]&:=&\int_\Sigma h\sqrt{\det
E}d^3x \label{cxii} 
\end{eqnarray} 
then one can work out the following
remarkable Poisson brackets: 
\begin{eqnarray} 
\left\{ {\cal H}^E[f]\,
,\, V[h]\right\} &=& 2\beta^2\kappa\int_\Sigma f\,h\,E^{ia}K^i_ad^3x
=\nonumber\\ &:=&  2\beta^2\kappa^2 T[fh] \nonumber\\ \left\{ T[f]\,
,\, E^{ia}(x)\right\} &=&+f(x)E^{ia}(x)\nonumber\\ \{ T[f]\; ,\,
K^{i}_a(x)\} &=& -f(x)K^i_a(x) \label{cxiii} \\ \{ T[f]\; ,\,
\Gamma^{i}_a(x)\} &=& \half E^{ib}\epsilon_{abc}q^{cd}\partial_df
\nonumber\\ \left\{ A^{(\beta )i}_a \, ,\, V[h]\right\} &=&
{\beta\kappa h\over 4\sqrt{\det
E}}\,\epsilon_{abc}\epsilon_{ijk}E^{ia}E^{jb} \nonumber 
\end{eqnarray}
I shall ignore possible problems with boundary terms, write simply
$V:=V[1],\ T:=T[1] ,\  H^E:={\cal H}^E[1]$, and summarize the above
as: 
\begin{eqnarray} 
\{  H^E,V\} =2\beta^2\kappa^2T  &;& 
\{T,E^{ia}\}=E^{ia}\nonumber\\ \{T,\Gamma^i_a\}=0 &;& \{T,K^i_a\}=-
K^i_a \label{cxv} 
\end{eqnarray} 
One use of these identities could be
to rewrite $\cal H$ in the (possibly more manageable) form:
\begin{eqnarray} 
\frac{\beta\kappa}{2} {\cal H}= \{ A^{(\beta)i}_a,V
\} \epsilon^{abc}F^{(\beta)i}_{bc}\quad\quad\quad\quad\quad
&&\label{cxvi}\\ - \frac{1+\beta^2}{\beta^2} \{ A^{(\beta)i}_a,V \}
\epsilon^{abc}\epsilon_{ijk}\{ A^{(\beta)j}_b,T \}\{ A^{(\beta)k}_c,T
\} &&\nonumber 
\end{eqnarray} 
This is the line followed by T. Thiemann
in \cite{Tb}, and claimed to be a runaway success.

A more fancy approach starts from the observation that the Hamiltonian
evolution induced by some $H$ on a function $f$ on phase space is
given by  the map: 
\begin{equation}
 W_H(t) \circ f:=f+t\{f,H\}
+{t^2\over 2!}\{\{f,H\},H\} +.. 
\end{equation} 
which preserves Poisson
brackets etc.; furthermore, from eq.(\ref{cxv}): 
\begin{eqnarray}
W_T(t)\circ E^{ia}&=&e^{-t}E^{ia}\nonumber\\ W_T(t)\circ
A^{(\beta)i}_a&=&\Gamma^i_a+\beta e^tK^i_a   \label{cxviii}
\end{eqnarray} 
One may say that this relation shows explicitly that
the value of $\beta$ does not matter, since it can be changed at will
(which is the point I am trying to make). More boldly, that it gives
us a way to go from the Barbero to the Ashtekar connection, setting
$t=i\pi /2$ if we had $\beta =1$ \cite{Ta,AII} . This  idea of a "Wick
rotation" is striking, but I find it very difficult to articulate. In
particular, I find difficult to understand how the transformation
brings about a change of (gauge) symmetry group. I nevertheless regard
this as the most promising clue to the construction of a satisfactory
quantum theory within this set of ideas. 
\section{ Spin nets.} 
To quantize the theory we may use the connection representation, in which
states are functionals of $A^{(\beta )i}_a(x)$, and 
\begin{equation}
E^{ia}\quad\to\quad \hat E^{ia} := {\beta\kappa\hbar\over
i}{\delta\over \delta A^{(\beta )i}_a}   \label{si} 
\end{equation} 
The important states turn out to be "spin net states", which are an
obvious generalization of Wilson loops.

\epsfbox{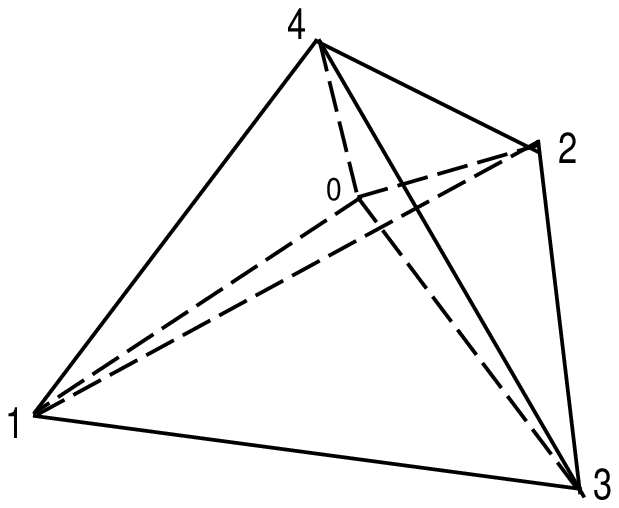}

Take for ex. the net in the picture, which has 5 vertices and 10
lines, and lives in a space with a connection $A^{(\beta )i}_a$;
assign to each line $\gamma_l$ an orientation, hence a "transporter"
$g_l:=P\exp\int_l\tau_iA^{(\beta )i}_ad\gamma_l^a$, and a spin
$s_l=0,\half ,1,\ldots$, so that $\gamma_l\to{\cal
D}^{s_l}_{mm'}(g_l)$; assign to each vertex $v$ an $SU(2)$ invariant
tensor $C^v_{m_a,..,m_d}$, and we shall have that: 
\begin{equation}
\psi_{\underline n}(g_1,...,g_{10})=\sum_{ \{ m \} }\prod_v
C^v_{...}\prod_l {\cal D}^{s_l}_{..}(g_l) 
\end{equation}
is gauge invariant. If $g_l\in SU(2)$, and we indicate by $dg_l$ the
Haar measure, these states form a (smallish) Hilbert space with the
scalar product: 
\begin{equation} 
<\psi_{{\underline
n}'}|\psi_{\underline n} > =\int\overline{\psi}_{{\underline
n}'}\;\psi_{\underline n} dg_1\ldots dg_{10} \label{sii}
\end{equation} 
Ashtekar and collaborators have discovered and
emphasized again and again (see e.g.\cite{AL} ) that if one considers
all possible nets, then the (very large) set of these states is {\it
dense} in the Hilbert space of gauge invariant functionals of 
$A^{(\beta )}$. In this sense eq.(\ref{sii}) induces a measure
$DA^{(\beta)}$ in this space.

The gauge invariant functionals of $A^{(\beta)}$ which might represent
physical states are invariant under diffeomorphism. Given one such
state $\Psi [A^{(\beta )}]$, for every spin net state 
$\psi_n[A^{(\beta )}]$ we can in principle calculate its "loop
transform": 
\begin{equation} 
\Psi ({\underline n}):= \int
DA^{(\beta)}\; \overline{\psi}_{\underline n}[A^{(\beta )}]\;\Psi
[A^{(\beta )}] \label{siii} 
\end{equation} 
that will represent the
same state, and is diffeomorphic invariant \underbar{if} it depends
only on the structure of the net and on our assigments of spins and
invariant tensors. This is the latest incarnation of the  "loop
representation" idea of Rovelli and Smolin\cite{RSa}. This form goes
particularly well with the idea of the "weave"\cite{ARS}:  that the
world that (we think) we know is likely to be described by states
$\Psi ({\underline n})$ with support on huge, immensely complicated
and fine meshed nets, in fact with  mesh sizes of the order of the
Planck length.

If we keep in mind the weave idea, it is quite sensible to look at net
states to find the spectrum of the operators that correspond to the
area of a surface $S$ and to the volume of a  region $R$, and to study
the operator form of the constraints.

Very briefly: if the surface $S$ intersects a subset  $\cal L$ of
lines of the net, does not touch the vertices, has no line lying on
it, carefully regularizing the operator ({\it first} taking the square
root, {\it then} removing the regulator), one finds\cite{AL,DPR} :
\begin{eqnarray} 
\hat A (S)\psi_{\underline n} &=& :\int_sd^2\sigma
\sqrt{n_an_b\hat E^{ia} \hat E^{ib} }\; :\;\psi_{\underline
n}\nonumber\\ &=& (\beta \hbar\kappa )\sum_{l\in {\cal
L}}\sqrt{s_l(s_l+1)}\;\psi_{\underline n} 
\end{eqnarray} 
If the region
$R$ contains a subset $\cal V$ of vertices of the net, one
finds\cite{DPR,L,Tc} by a similar procedure that a suitably
regularized volume operator $\hat V(R)$ mixes the states
$\psi_{\underline n} $ with coefficients proportional to
$(\beta\hbar\kappa )^{3\over 2}$. This operator can be
diagonalized\cite{DPR} , and has a complicated, but discrete spectrum.
Before going further, it is important  to notice that nothing works
for the theory based on the Ashtekar connection $\beta =i$: all
operators have the wrong hermiticity; on this, see later.

However, I claim that the discrete spectra one gets for areas and
volume \underbar{cannot} at this stage be interpreted as evidence for
a discrete structure of space, because of the arbitrariness of
$\beta$; we are faced with a "$\beta$ crisis". Unless we find some
good reason to fix $\beta$, the commutation relations eq.(\ref{cix})
will be unable to fix the scale of the theory. My feeling is that this
requires a group larger than $SU(2)$; a bit like in good old  current
algebra. A larger group may come from the need to implement the "Wick
rotation", whether interpreted as a passage from the Euclidean to the
Minkowskian or from the Barbero to the Ashtekar connection I do not
know.

Alternatively, it may be that looking at the Wheeler De Witt equation
$\hat{\cal H}\cdot\Psi =0$, following Thiemann's work \cite{Tb} , we
shall find that solutions exist only for particular values of $\beta$.
Given the way in which eq.(\ref{cxvi})  depends on $\beta$, this is
actually rather likely; but then, it m     ay be just another way of
saying the same thing. At the same time, if it were true it would be
splendid: because we would have derived the discreteness of space from
dynamics, and not from a kinematic fiddle. \epsfbox{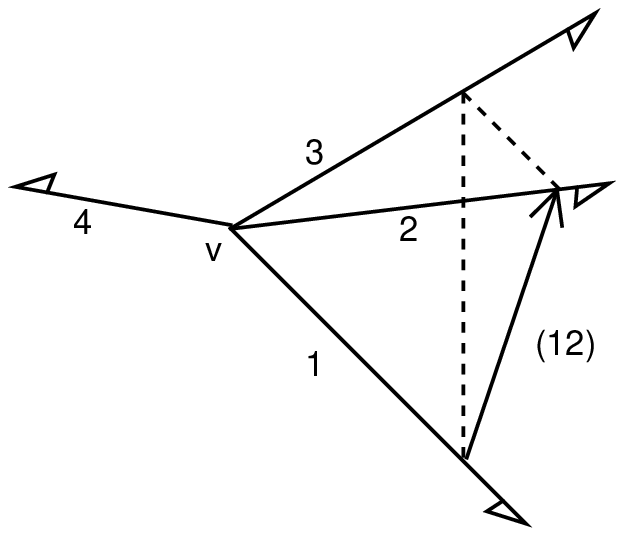}

Thiemann's work \cite{Tb} on the Hamiltonian constraint is the latest
thing. It is a massive and difficult piece of work, so that my
comments, based on a very limited understanding, are obviously very
tentative. The main effort is aimed at defining a satisfactory quantum
operator corresponding to ${\cal H}^E[f]$, eq.(\ref{cxii}), from
which, in view of eq.(\ref{cxiii},\ref{cxv}) the rest follows (more or
less).  A regularization is obtained noticing that for a vertex $v$ of
a net and a choice of three (outgoing) lines, say $1,\ 2,\ 3$, adding
lines $(IJ)$ (see the figure), in the na\"\i ve limit $g_I\approx
1+A_a\Delta_I^a+\ldots$ one has: 
\begin{eqnarray} 
{4\over
3\beta\kappa} \sum_{I,J,K}\epsilon^{IJK}\Tr
(\,g_Ig_{IJ}g_J^{-1}\;g_K\,\{g_K^{-1}, V \})&&\nonumber\\ \approx
{1\over\sqrt{\det E}} \epsilon_{ijk}E^{ia}E^{jb}F^{(\beta
)k}_{ab}\;{\cal V}_{(123)}  && \label{svi} 
\end{eqnarray} 
where ${\cal
V}_{(123)}:={1\over 6}\epsilon_{abc}\Delta_1^a\Delta_2^b\Delta_3^c$ is
the coordinate volume of the tetrahedron shown. The action of the
quantum $\hat{\cal H}^E[f]$ on net states is defined summing over all
vertices and all choices of three  lines the corresponding operator:
\begin{eqnarray} 
\hat{\cal H}^E[f]\cdot \psi_{\underline n}=
\sum_{v,\{IJK\} }{4\over 3i\beta\hbar\kappa}f(v)\cdot
\quad\quad\quad\quad\quad &&\nonumber\\
\cdot\sum_{I,J,K}\epsilon^{IJK}\Tr
(\,g_Ig_{IJ}g_J^{-1}\;g_K\,[g_K^{-1}, \hat V ]) \cdot\psi_{\underline
n}  && \label{svii} 
\end{eqnarray} 
This is only the beginning, the
definition is sharpened and modified several times en route. I
understand that this summer it has been the object of intensive
discussions by lots of specialists at the ESI workshop, and no doubt
you will hear more and in more detail about it.

\section{ Discretizing.} 
The pattern I have described so far occurs
again if you try and discretize the theory, except that there are more
variants: from the variable mesh discretizations of the groups that
solve problems in the classical theory numerically (an industry in
rapid expansion), to emulatations of lattice gauge theory or to the
theory of random triangulations. I shall restrict myself drastically,
and try to  keep close in spirit to the original Regge idea.

Divide space in tetrahedra  with a flat inside:  the natural variables
corresponding to the $E^{ia}$ would be the areas of the triangles,
oriented outwards with respect to a frame local to the tetrahedron:
\begin{equation} 
S:=\tau_iS^i:=\tau_i E^{ia}\epsilon_{abc}\half
(x_1x_2+x_2x_3+x_3x_1)^{bc} \label{ri} 
\end{equation} 
The area square of a triangle will be $S^iS^i$, and of course 
tetrahedra must close;
for the tetrahedron $A=(1234)$, labeling each triangle with the number
of the vertex opposite, this gives the constraint: 
\begin{equation}
S^i_1+S^i_2+S^i_3+S^i_4=0 \label{ria} 
\end{equation} 
The volume square
of the tetrahedron will be : 
\begin{eqnarray*}
V_A^2=\qquad\qquad\qquad\qquad\qquad\qquad &&\\ {\epsilon_{ijk}\over
18} (S^i_2S^j_3S^k_4+S^i_1S^j_4S^k_3 +S^i_2S^j_4S^k_1+S^i_3S^j_2S^k_1)
&& 
\end{eqnarray*} 
The variables $S$ must be paired with variables $g$
that link the frame attached to tetrahedron $A$ to the one of its
neighbour $B$ across the triangle, with the basic condition that the
triangle looks the same from both sides: 
\begin{equation} 
S_B = -g^{-1}_{AB} S_A g_{AB}  \label{rii} 
\end{equation}

\epsfbox{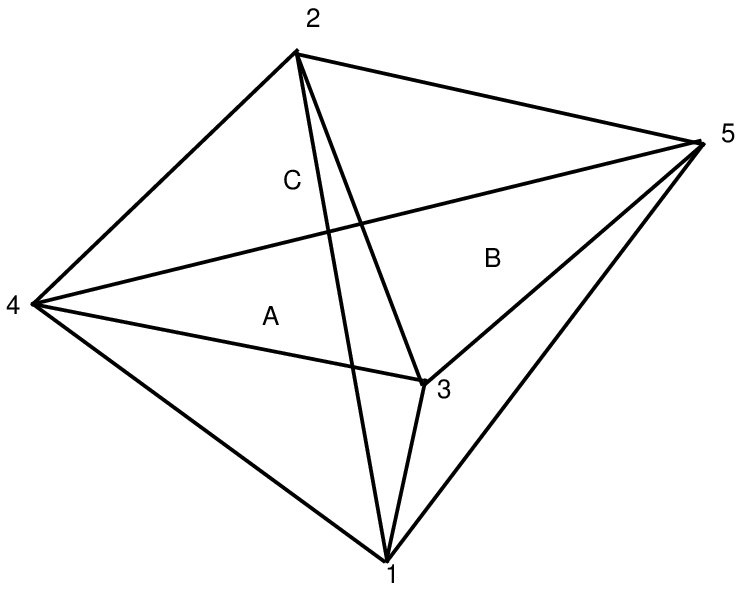}

If $g$ is a rotation, we will discover the local geometry of our space
going round an edge, from tetrahedron to tetrahedron, until we get
back: then if we are not in the same frame we started from, we
conclude that space is curved. Changes of the  frames, i.e. local
rotations, are "gauge" transformations, which translate eq.(\ref{cvi})
to: 
\begin{eqnarray} 
S_A\to g_AS_Ag_A^{-1} &;&   S_B\to
g_BS_B^{-1}g_B\nonumber\\ g_{AB}&\to& g_Ag_{AB}g_B^{-1}   \label{riii}
\end{eqnarray} 
An explicit parametrization for the triangle $(123)$
would be : 
\begin{eqnarray} 
S_A :=u_A\cdot\tau_3s\cdot u_A^{-1}  &; &
S_B:=u_B\cdot\tau_3s\cdot u_B^{-1}\nonumber\\ {\rm with:}\
u_A=e^{\alpha \tau_3}e^{\beta \tau_2} & ;& u_B:= e^{\gamma
\tau_3}e^{\delta \tau_2}\nonumber\\ g_{AB}:=u_A e^{\Phi
\tau_3}2\tau_2u_B^{-1}&& \label{riv} 
\end{eqnarray} 
with $s>0$; the angle $\Phi$ is arbitrary at this stage.

Suppose that we are given the lengths of all edges, like in Regge's
original scheme for "general relativity without coordinates": then we
are in a Riemannian space, and we have conditions for the triangles
that share an edge; for the edge (12) they read: 
\begin{equation}
[[S_{3B},S_{5B}]\, ,\,g^{-1}_{AB}\;[S_{3A},S_{4A}]\;g_{AB}]=0 
\label{rv} 
\end{equation} 
I omit the details: the point is that from
these relations we can calculate all the angles $\Phi$, and for the
curvature around the edge $(12)$ we find: 
\begin{eqnarray} 
R_A:= e^{F_A}=\quad\quad\quad\quad\quad\quad\quad &&\label{rvi}\\
=u_{4A}e^{\phi_A\tau_3}\;e^{(\theta_A+\theta_B+\theta_C-2\pi)\tau_2}
\,2\tau_2\;e^{-\phi_A\tau_3}u_{4A}^{-1} &&\nonumber 
\end{eqnarray}
Curvature is essentially the "defect angle" in the rotation; the
connection is Levi--Civita.

But what can we make of the $K^i_a$? To make the connection "dynamic",
we must give up eq.(\ref{rv}), and therefore (\ref{rvi}), and leave
the angles $\Phi$ arbitrary, because that is the only place where the
extrinsic curvature can go.  Omitting details again, by various
plausibility arguments, and insisting that in the limit eq.(\ref{cix})
is recovered, one finds for the Liouville form $\Theta$ (the $\int
pdq$ bit of the action) and for the Poisson brackets: 
\begin{eqnarray} 
\Theta &=&-{1\over\beta\kappa}\sum_T\sum_{t\in T}\Tr
\big(S_t\delta g_t g^{-1}_t\big)_T\nonumber\\ \{
S^i,S^j\}&=&\beta\kappa\,\epsilon_{ijk}S^k\ ;\ \{
g,S^i\}=\beta\kappa\, \tau_i\,g \nonumber\\ \{ g^{(1)},g^{(2)}\}&=&0
\label{rvii} 
\end{eqnarray} 
In the explicit parametrization
eq.(\ref{riv}) each triangle contributes to the Liouville form:
\begin{eqnarray} 
-{2\over\beta\kappa}\Tr (S\,\delta g\,
g^{-1})=\qquad\qquad\qquad\qquad && \nonumber\\
={1\over\beta\kappa}(s\,\delta\Phi +s\cos\beta\,\delta\alpha +
s\cos\delta\,\delta\gamma )  && \label{rviii} 
\end{eqnarray} 
I have
not tried to write the constraints for real $\beta$. The idea of
squeezing in dynamics by fiddling $\Phi$ may appear ugly and
artificial, but it is the sense of the "Barbero connection". The
Ashtekar choice makes much more sense: one replaces the $\Phi$ we
calculated above with a complex variable: 
\begin{equation}
\Phi\quad\to\quad\Phi +i\zeta   \label{rix} 
\end{equation} 
so that $g_{AB}$ is promoted to a Lorentz transformation, with rapidity
$\zeta$, that links the different inertial frames attached to
tetrahedra $A$ and $B$. One may say that this is physically well
motivated, and follows straight from the principle of equivalence.
However, extending the gauge transformations eq.(\ref{riii}) from
$SU(2)$ to $SL(2,\Complexes )$ inevitably creates complications:  we
have to impose explicitly that within a tetrahedron $S^i_IS^i_J$ is
real and positive definite, and eq.(\ref{rv}) has to be modified. On
the other hand we know that the Hamiltonian constraint simplifies to
the ${\cal H}^E$ form, for which it is not difficult  to guess 
discrete  versions, for ex.:
\begin{equation} 
\sum_{I<J\in T}{1\over V_T}\Tr ([S_I,S_J] F_{IJ})_T\;
\approx\;  0  \label{rx} 
\end{equation} 
in words: for each
tetrahedron, the sum of the lengths of the edges times the
corresponding defect angle must vanish. This comes about because for
the edge $(12)$ of A we can see that \begin{equation} l^i_{(12)}:=
{2\over 3V_A}\epsilon_{ijk}S^j_4S^k_3\ \approx\ e^i_a(x_2^a-x^1_a) 
\label{rxi} \end{equation} In practice  to make any use of
eq.(\ref{rx}), which is certainly not the only expression possible, we
must approximate $F$ with some function of $R$, e.g.$F\approx
(R-R^{-1})/2$. The same or a similar expression can be used for the
${\cal H}^E$ part of the Hamiltonian constraint for real $\beta$. It
is unfortunate that, as far as I can see, the regularized form used by
Thiemann is just about the least natural in this scheme. This is
because, at the first step one finds, for ex.: 
\begin{eqnarray*} 
g_4\{ g_4^{-1},V_A\} ={\beta\kappa\tau_i\epsilon_{ijk} \over 36V}
(S_2^jS_3^k+S_3^jS_1^k+S_1^jS_2^k)&&\\ ={\beta\kappa\over
24}\tau_i(l^i_{(41)}+l^i_{(42)} +l^i_{(43)}) && 
\end{eqnarray*} 
a rather uncooperative expression; or, said more generally, because
Thiemann's tetrahedra live in a lattice  dual to the Regge
lattice\footnote{ 
notice that in eq.(\ref{svi}) the volume operator
$\hat V$ acts on the vertices of the net, that correspond to Regge
tetrahedra, while ${\cal V}_{(123)}$ is the volume of a tetrahedron in
the dual lattice.},
 in which to each of our triangles corresponds  a
line, and to each tetrahedron a (4--valent) vertex.

Now for quantization. At first sight, from what I said about nets, it
would appear that discretization, and Regge calculus in particular,
offers the perfect tool; in fact, I became interested in Regge
calculus because I was trying with spin nets. Spin nets naturally live
in the dual of the Regge lattice. One can immediately envisage putting
on a (small) computer a finite, simple lattice like a five tetrahedra
division of $S^3$. In a real $SU(2)$ formulation the quantization of
areas follows directly from eq.(\ref{rviii}): the area variable $s$ is
conjugate to an angle $\Phi$. This is just the way 't Hooft argued in
his discretized 2+1 gravity\cite{tho} , but with a very refined
argument to Wick-rotate the theory. However, the direct use of the
Ashtekar connection for quantization is made impossible by a muddle
over the measure. I shall explain this in detail because I still hope
that the muddle has a simple solution that I cannot see because of
some selective blindness.

One can see what the problem is quite simply by writing the quantum
version of eq.(\ref{rvii}) for $\beta=i$: 
\begin{equation}
[\hat{S}^i,\hat{S}^j ] = -\kappa\,\epsilon_{ijk}\hat{S}^k\ ;\quad
[\hat{g},\hat{S}^i] = -\kappa\,\tau_i\,\hat{g}   \nonumber
\end{equation}
 \begin{equation} 
 [\hat{g}^{(1)},\hat{g}^{(2)}] = 0
\end{equation} 
so that if $\psi =\psi (\{ g\}),\ \hat{S}^i= -\kappa
\hat{T}^i_L$, the (holomorphic) generator of the left--regular
representation of $SL(2,\Complexes )$. The problem is that now there
is no way to juggle the measure to make $\hat S^i$ hermitean; worse,
$\hat S^2$ is negative, with eigenvalues $-j(j+1)$. Of course one did
expect troubles: there are always troubles with non--compact groups,
e.g. in 2+1 gravity\cite{ALo}, and even linearized gravity turns out
to be quite tricky\cite{ARSb}; in QED the scalar product is well
defined only for gauge invariant states. So the cure should come from
considering the Hamiltonian constraint; in an interesting analysis of
the linearized case\cite{RPW} the suggestion is that one should gauge
fix it. In our case, the idea of "Wick rotation" seems much more
attractive, but notice that in any case it depends on solving the
Hamiltonian constraint.

The idea\cite{Ta} is that one can apply to this transformation the
formalism of "coherent state transformations" developed by B.C.
Hall\cite{BCH} , suitably generalized. These are isometries between
Hilbert spaces on e.g. $SU(2)$ and on its complexification
$SL(2,\Complexes )$; in the simplest case one defines heat kernels on
the two groups by: 
\begin{eqnarray} 
{\partial\rho_t\over \partial
t}+\half\hat{\bf J}^2\rho_t =0\ &,&\ \rho_0=\delta (x)\nonumber\\
{\partial\mu_t\over \partial t}+\quart (\hat{\bf J}^2+\hat{\bf
K}^2)\mu_t =0\ &,&\ \mu_0=\delta (g)   \label{rxx} 
\end{eqnarray}
(notice that $(\hat{\bf J}^2+\hat{\bf K}^2)$ is an elliptic operator,
but not the Casimir). Then the transformation $B_t$: \begin{equation}
f \to (B_tf)(g):=\int_{SU(2)}f(x)\rho_t(x^{-1}g)dx \end{equation} maps
functions on $SU(2)$ which are $L^2$ with the measure $\rho_tdx$ to
functions on $SL(2,\Complexes )$ which are holomorphic and $L^2$ with
the measure $\mu_tdg$. The map can be proved to be invertible and
isometric, i.e. 
\begin{eqnarray*} 
\int_{SL(2,C)} \overline{(B_tf)(g)}\
(B_th)(g)\;\mu_t(g)dg=&&\\
=\int_{SU(2)}\overline{f(x)}\;h(x)\rho_t(x)dx  && 
\end{eqnarray*} 
This seems to be just what we need, but the connection with the "Wick
rotation" is still to be understood.

\end{document}